\documentclass[aps,prd,twocolumn,superscriptaddress,amsmath,amssymb,floatfix]{revtex4-2}

\usepackage{graphicx}% Include figure files
\usepackage{dcolumn}

\usepackage{appendix}
\usepackage[hidelinks]{hyperref}
\usepackage{color}
\usepackage{lineno}
\usepackage{comment}

\usepackage[defaultcolor=red,commandnameprefix=ifneeded,final]{changes}
\begin{document}

%\linenumbers
%\preprint{}

%\title{Prospects for Joint Detection of Gravitational Waves with Counterpart Gamma-Ray Bursts Detected by the HADAR Experiment}
\title{Two-component diffuse Galactic gamma-ray emission revealed with Fermi-LAT}



\author{Qi-Ling Chen}
\affiliation{College of Physics, Sichuan University, Chengdu 610064, P.R.China}
\affiliation{Key Laboratory of Particle Astrophysics, Institute of High Energy Physics, Chinese Academy of Sciences, Beijing 100049, P.R.China}
 
\author{Qiang Yuan}
\email{yuanq@pmo.ac.cn}
\affiliation{Key Laboratory of Dark Matter and Space Astronomy, Purple Mountain Observatory, Chinese Academy of Sciences, Nanjing 210023, P.R.China}
\affiliation{School of Astronomy and Space Science, University of Science and Technology of China, Hefei, Anhui 230026, P.R.China}
 
\author{Yi-Qing Guo}
\email{guoyq@ihep.ac.cn}
\affiliation{Key Laboratory of Particle Astrophysics, Institute of High Energy Physics, Chinese Academy of Sciences, Beijing 100049, P.R.China}
\affiliation{University of Chinese Academy of Sciences, 19 A Yuquan Rd, Shijingshan District, Beijing 100049, P.R.China}
 
\author{Ming-Ming Kang}
\email{kangmm@ihep.ac.cn}
\affiliation{College of Physics, Sichuan University, Chengdu 610064, P.R.China}

\author{Chao-Wen Yang}
\affiliation{College of Physics, Sichuan University, Chengdu 610064, P.R.China}

%\author{}
%\thanks{}
%\affiliation{}


\date{\today}

\begin{abstract}

The enigma of cosmic ray origin and propagation stands as a key question in particle astrophysics. The precise spatial and spectral measurements of diffuse Galactic gamma-ray emission provide new avenues for unraveling this mystery. Based on 16 years of Fermi-LAT observations, we find that the diffuse gamma-ray spectral shapes are nearly identical for low energies (below a few GeV) but show significant dispersion at high energies (above a few GeV) across the Galactic disk. We further show that the diffuse emission can be decomposed into two components, a universal spectral component dominating at low energies which is consistent with the expectation from interactions of background cosmic rays and the interstellar matter, and a spatially variant component dominating at high energies which is likely due to local accelerators. These findings suggest that there is dual-origin of the Galactic diffuse emission, including the ``cosmic ray sea'' from efficient propagation of particles and the ``cosmic ray islands'' from inefficient propagation of particles, and thus shed new light on the understanding of the propagation models of Galactic cosmic rays.

\end{abstract}

%\keywords{}

\maketitle

%\section{\label{sec:introduction}Introduction}

Galactic cosmic rays (CRs) are believed to be accelerated by extreme accelerators in the Milky Way, such as the remnants of supernova explosions or jets from black hole accreting systems. These energetic particles are then injected into the Milky Way, propagate diffusively and interact with the interstellar medium (ISM) and radiation field (ISRF), producing secondary particles and radiation. Measurements of the spectra of primary and secondary CRs, as well as diffuse $\gamma$ rays, verify that, at the zeroth order approximation, this conventional paradigm can hold (at least for the low-energy band with $E<$~TeV/n) \cite{Strong:2007nh}.

%*****************************Figure 1***********************************
\begin{figure*}[thb]
\centering
\includegraphics[width=0.8\linewidth]{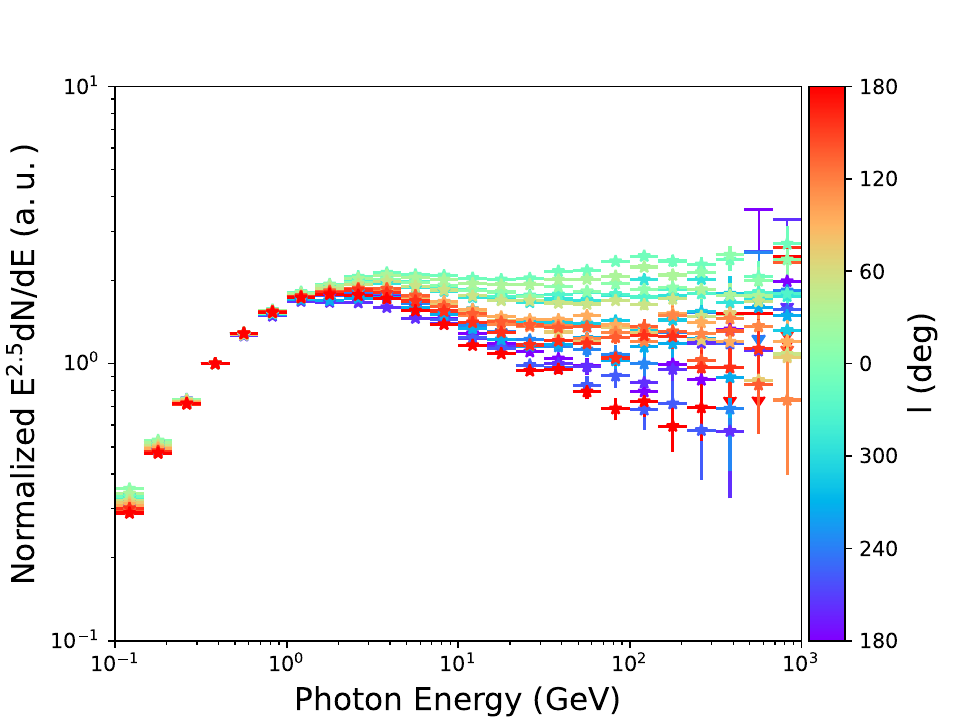}
\caption{SEDs of diffuse Galactic $\gamma$-ray emission from 18 segments of the Galactic plane.
All the fluxes are normalized at 0.6 GeV. Different colors represent different Galactic longitude bins.
}
\label{fig:sed_all}
\end{figure*}
%*************************************************************************

Recent new observations indicate that this conventional paradigm needs to be refined or revised.
Direct measurements of spectra of primary CRs revealed spectral hardenings around hundreds of GV
and subsequent softenings around $O(10)$ TV \cite{PAMELA:2011mvy,AMS:2015tnn,Yoon:2017qjx,Atkin:2018wsp,AMS:2021nhj,DAMPE:2019gys,Alemanno:2021gpb,CALET:2022vro,CALET:2023nif}. Combined with the correlated structures in the large-scale anisotropies,
a nearby source contribution seems to be a natural explanation of these features in both the
spectra and anisotropies \cite{Savchenko:2015dha,Ahlers:2016njd,Liu:2018fjy}.
On the other hand, observations of secondary spectra of CRs show more prominent hardenings than
the primary particles \cite{AMS:2018tbl,DAMPE:2022jgy,DAMPE:2024qwc}. Furthermore, measurements
of diffuse $\gamma$-ray emission in the very to ultra high-energy bands
\cite{Milagro:2005xqq,TibetASgamma:2021tpz,LHAASO:2023gne,LHAASO:2024lnz,HAWC:2023wdq} 
and also the diffuse Galactic neutrino emission \cite{IceCube:2023ame} also show excesses compared 
with the conventional CR propagation model predictions. These secondary CRs and radiation excesses 
may indicate that more grammages of high energy particles experience than expected in the conventional
transport scenario, possibly due to confinement of particles around the acceleration
sources \cite{Zhang:2021xri,Sun:2023ibg,Yang:2024igs}. The above primary and secondary spectral 
features may be unified in the sense that, the nearby source close to the Earth may exist generally 
in the Milky Way, leaving imprints on the secondary nuclei and diffuse emission.

Here we report the analysis of the diffuse $\gamma$-ray emission from the Galactic plane with 16
years of Fermi-LAT data, to test the new propagation and interaction scenario of Galactic CRs.
We divide the Galactic plane (with $-10^{\circ}<b<10^{\circ}$) into 18 regions, each spanning a 
longitude range of $20^{\circ}$. The resolved sources in the 4th Fermi-LAT 
catalog \cite{Fermi-LAT:2019yla,Fermi-LAT:2022byn} and the isotropic diffuse 
emission have been subtracted, and the spectra and spatial distributions of the Galactic diffuse
component are derived. See the Methods for more details of the data analysis. 
The obtained spectral energy distribution (SEDs) of the 18 regions, normalized at 0.6 GeV, are
shown in Figure \ref{fig:sed_all}. One can see that at low-energies ($E\lesssim$~GeV), the SEDs
of all these regions are highly consistent with each other, suggesting a uniform origin of 
the emission. However, for energies above a few GeV, significant variations of the fluxes and
spectral shapes are present. We can also find a trend that the high energy spectra are harder
in the Galactic center regions than in the anti-center regions. A spectral index close to $2.5$
in the inner Galaxy region is also consistent with previous results by the Fermi-LAT 
collaboration \cite{Fermi-LAT:2012edv}.

%*****************************Figure 2***********************************
\begin{figure*}[thb]
\centering
\includegraphics[width=0.48\linewidth]{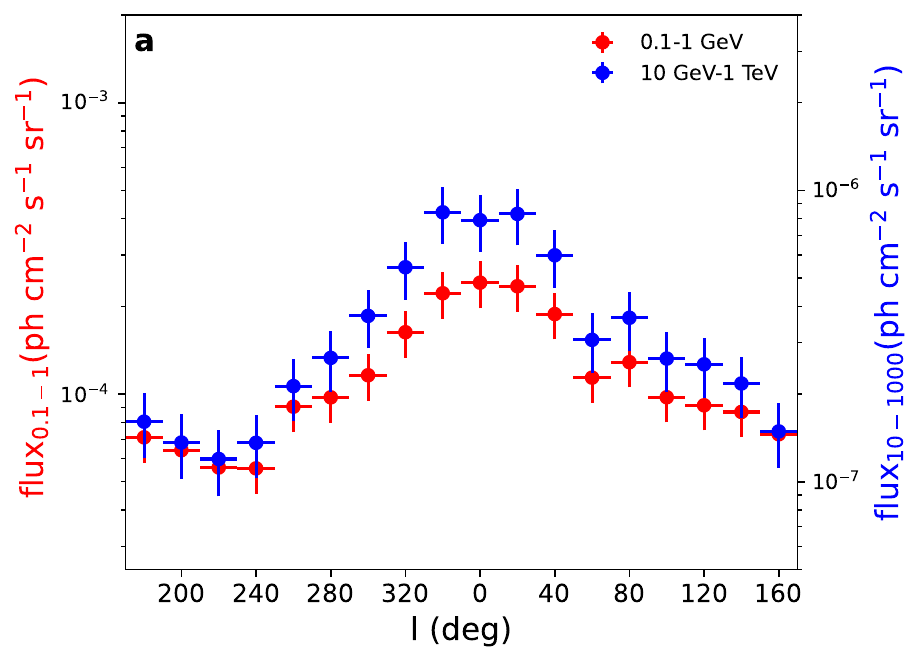}
\includegraphics[width=0.48\linewidth]{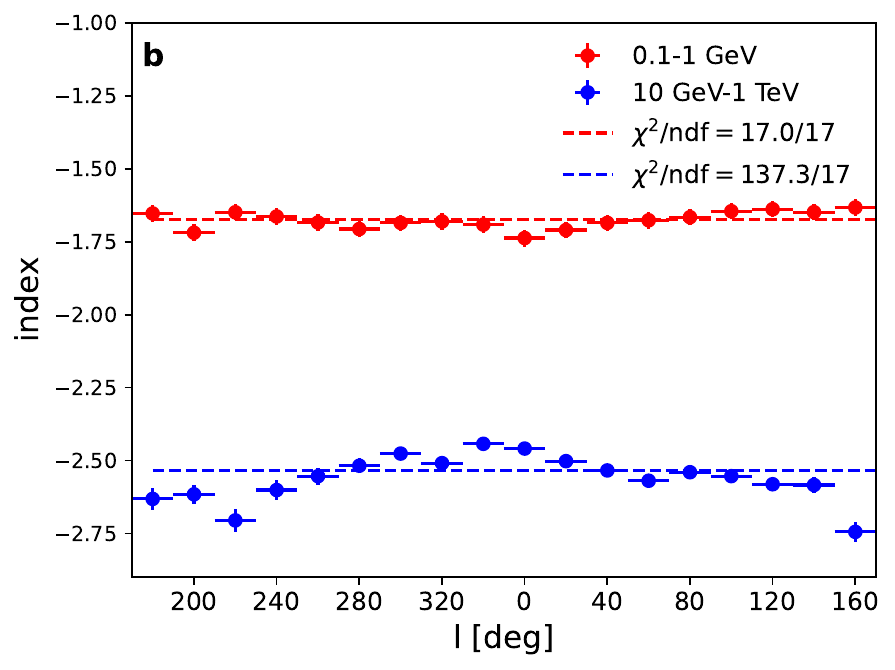}
\caption{The fluxes (a) and photon indices (b) in different sky regions.
Blue color is for $0.1-1$ GeV band, and red is for $10-1000$ GeV band. In panel b, 
dashed lines correspond to fitting results of the spectral indices with a constant. 
}
\label{fig:flux_index}
\end{figure*}
%*************************************************************************

To better see the variations of the emission accross the Galactic plane, we show in Figure
\ref{fig:flux_index} the fluxes (panel a) and power-law indices (panel b) obtained through fitting
the SEDs in the $0.1-1$ GeV and $10-1000$ GeV bands, respectively. The fluxes in both energy bands 
show clear spatial variations in different regions, with higher fluxes in the inner Galaxy regions
and lower fluxes in the outer regions. However, the high energy fluxes show more prominent variations 
than the low energy ones. The spectral indices are distinct in these two energy bands, as shown in
panel b. In the low energy band, the indices are almost constant, $\sim 1.67$, across the Galactic 
plane. The fitting with a constant yields a $\chi^2$ over degree-of-freedom (dof) of $17.0/17$. 
For the high energy part, the spectral indices fluctuate significantly. A constant fitting gives
$\chi^2/{\rm dof}=137.3/17$, indicating that the data deviate from the constant assumption at a
significance of $9.4\sigma$.

%The flux and index of each region are shown in the Fig.~\ref{fig:flux_index}. In Fig~\ref{fig:flux_index} {\bf a}, orange dots and blue dots represent the 2-20 GeV and 0.1-1 TeV flux respectively. In {\bf b}, the colors are the same. From this figure, we also find that, the high energy flux doesn't consistent with the low energy flux.  Compared to low-energy flux, high-energy radiation exhibits greater fluctuations. As shown in the {\bf b} in the figure, the index at low energy has more similarity than high energy.

%*****************************Figure 3***********************************
\begin{figure*}[thb]
\centering
\includegraphics[width=0.8\linewidth]{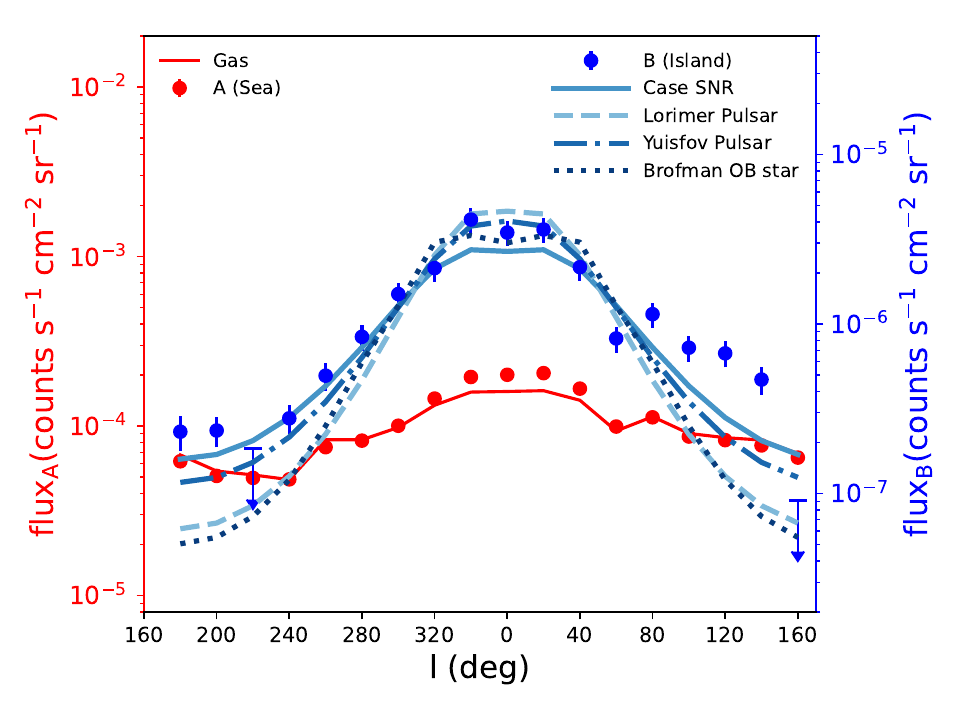}
\caption{Tow component flux which is integrate of spectrum. Component A is derived from the locally observed CR spectrum after correction of the solar modulation, and component B is the additional cutoff power-law spectrum. The red line represents the gas distributions. The blue lines represent the source distributions for supernova remnants \cite{Case:1998qg}, pulsars \cite{Lorimer:2006qs,Yusifov:2004fr}, and OB stars \cite{Bronfman:2000tw}. Component A is well tracked by the gas distribution, and component B is consistent with the source distributions.}
\label{fig:flux_glon}
\end{figure*}
%************************************************************************

%Figure~S\ref{fig:sed_all_mod} incorporates the proton spectrum for comparison on the basis of Figure~\ref{fig:sed_all}. The data of proton spectrum comes from the Voyager, AMS-02, DAMPE, correspond to the red, magenta, blue dots respectively. Hollow dots represent the AMS-02 data considering the solar modulation.  The proton spectrum is normalized at 2 GeV. Since the cross-section has an energy dependence with an exponent of -0.1, the proton spectrum is additionally multiplied by an exponent of 0.1. Furthermore, because the energy loss for gamma-ray production is approximately $90 \%$, the proton energy is multiplied by 10. With this figure, we find that the proton spectrum has consistency with gamma spectrum below the 2 GeV, for high energy, they have difference. 
 
%Then We use the spectrum derived from proton spectrum to fit the data. The details can be seen in \ref{xxx}. In the figure, red line represent the spectrum from cosmic ray. Blue line represent the PL spectrum. The black dot is the SED of different regions. These results demonstrate that the energy spectrum can be interpreted as a superposition of the cosmic-ray sea spectrum and a power-law spectrum. Below 1 GeV, the data has a little excess over theory, maybe there is the contribution of bremsstrahlung. 

The spectral variations indicate that the low-energy and high-energy parts of the diffuse emission may 
have different origin. We decompose the emission into two components: the A component which dominates 
in the low-energy band and the B component which is more significant at high energies. We further 
note that the diffuse emission in the low-energy band is close to the expected $\gamma$-ray spectrum 
from hadronic interactions between the measured proton fluxes and the ISM, as shown in  
Figure~\ref{fig:proton_gamma}. Therefore we fix the spectrum of component A to be the expected one
from locally measured proton spectrum, leaving only the flux normalizations free to be fitted.
For component B, we also assume it has a hadronic origin, and parameterize the corresponding
proton spectrum as a power-law with low-energy cutoff. The low-energy cutoff is phenomenologically
employed to avoid over-shooting the data at low energies. The spectral index and cutoff energy
of component B are assumed to be the same in the 18 regions, and the flux normalizations are 
free to vary. We fit the spectra in the 18 regions for $E>0.5$ GeV since at low energies there
should be contribution from CR electrons and positrons. The derived fluxes of the two components, 
integrated in an energy band of $0.1-1000$ GeV, are shown in Figure~\ref{fig:flux_glon}, and the
best-fitting spectra compared with the measurements of the 18 regions are given in 
Figure~\ref{fig:flux_componment}. For comparison, we overplot in Figure~\ref{fig:flux_glon} the
gas distribution from the PLANCK dust opacity measurements \cite{Planck:2016frx} and the distributions
of several candidate CR source populations including supernova remnants \cite{Case:1998qg}, 
pulsars \cite{Lorimer:2006qs,Yusifov:2004fr}, and OB stars \cite{Bronfman:2000tw}. It is very
interesting to find that the longitude distribution of component A is close to that of the gas
distribution, while component B is closer to those of CR sources. For the four types of source
distributions, the pulsars given in Ref.~\cite{Yusifov:2004fr} and supernova remnants seem to match 
with the data better than the others. 

The findings obtained in this work thus point to a paradigm with dual-origin of the Galactic
diffuse $\gamma$-ray emission. Component A is more smoothly distributed in the Galaxy with a 
universal spectral shape, which may correspond to a ``background sea'' due to CRs experiencing
adequate propagation. Component B has a harder spectrum and is more concentrated to the inner
Galaxy region with a possible association with the source distribution, which may be due to
recently accelerated particles not adequately propagate throughout the Galaxy. In contrast to 
the low-energy ``sea'' component, we define component B as ``CR island''. This dual-component of 
the diffuse emission reflects the natural expectation of the time-dependence of the production 
and propagation of CRs. For particles freshly accelerated, they may get trapped in the source 
region and interact with the ISM surrounding the sources. With the increase of time, those 
particles would diffuse out and interact with the diffuse ISM throughout the Galaxy. For those 
sources compact and bright enough, they may be identified (as pointlike or extended source) and 
subtracted in the analysis. The rest sources form an unresolved population of low surface brightness 
sources and a truly diffuse component, which give the measured ``diffuse'' emission of this work.

Our results are likely correlated with the excesses of the diffuse $\gamma$-ray emission above TeV 
revealed by several experiments \cite{TibetASgamma:2021tpz,LHAASO:2023gne,LHAASO:2024lnz,HAWC:2023wdq}.
Compared with the model prediction from the CR sea, the data show excesses from several GeV to 
$\sim60$ TeV which could be explained by an exponential cutoff power-law component with spatial 
distribution being consistent with the assumed source distribution \cite{Zhang:2023ajh}. One of the 
most natural explanations is the unresolved sources \cite{Linden:2017blp,Yan:2023hpt,Vecchiotti:2021yzk}. 
Remarkably extended sources in the very high energy $\gamma$ rays have been 
detected \cite{LHAASO:2023uhj}, which may just reflect the fact that they are due to particles
not propagating far away from the acceleration sites. Considering that sources like the Cygnus
bubble may exist in general in the Galaxy, the diffuse $\gamma$-ray excesses can be well accounted
for \cite{Zhang:2021xri,Sun:2023ibg,Yang:2024igs}.

The two-component Galactic diffuse $\gamma$-ray emission offers new insights in understanding 
the long-standing puzzle of the origin and propagation of Galactic CRs. It indicates that the
traditional steady-state calculation of the propagation of CRs and the production of secondary
particles and radiation needs to be refined to include the time evolution \cite{Marinos:2024rcg}. 
Further tests of this paradigm may include precise measurements of the diffuse emission above TeV 
energies without source masks, as well as multi-messenger observations of CRs and neutrinos.

\begin{acknowledgments}
This work is supported by the National Natural Science Foundation of China (No. 12220101003, 
No. 12333006, No. 12263004) and the Project for Young Scientists in Basic Research of Chinese 
Academy of Sciences (No. YSBR-061).
\end{acknowledgments}

\appendix
\section{Fermi-LAT data analysis}
We select 16 years of the Fermi-LAT data, from August 4, 2008 to August 4, 2024 in the analysis. 
The reconstruction version of the data is {\tt P8R3\_SOURCE\_V3}. To reduce the effect from the
Earth atmosphere, we exclude the events with zenith angles larger than $90^{\circ}$. The energy
range of $0.1-1000$ GeV is selected. The data are binned into 6 logarithmically even energy bins 
per decade, and $0.1^{\circ}\times 0.1^{\circ}$ angular bins. We use {\tt Fermipy} v1.4.0 to 
analyze the data. Detected sources are modelled based on the 4FGL-DR4 catalog \cite{Fermi-LAT:2022byn}. 
The region of interest (ROI) of the analysis is $|b| < 10^{\circ}$, and the whole Galactic plane is
further divided into 18 sub-regions, each with $20^{\circ}$ span in longitudes. The source map
has a size of $30^{\circ}\times 30^{\circ}$ covering each sub-region to reduce the edge effect 
due to the point spread function (PSF). The Galactic background model used is {\tt gll\_iem\_v07.fits}, 
and the isotropic diffuse background used is {\tt iso\_P8R3\_SOURCE\_V3.txt}. For each ROI, we re-fit 
the normalization parameters of all the sources and the diffuse backgrounds. After the fitting, 
we subtract the detected sources and the isotropic background from the data, and ascribe the 
rest as Galactic diffuse emission. The Galactic diffuse fluxes are then computed with the residual
number of photons divided by the energy interval of each bin, the mean exposure, and the solid angle 
of the ROI. Due to the large PSF for energies below 500 MeV, we apply an energy-dependent PSF correction
using the instrument response model provided in {\tt gll\_iem\_v07.fits}, i.e., to estimate the
fractions of photons entering and leaking out of the ROI for given spatial template. 
%The correction was computed by dividing the fluxes within a $20^{\circ}\times 20^{\circ}$ region by the convolution of the fluxes with a Gaussian kernel whose width corresponds to the PSF. For the convolution, we evaluated the fluxes over a larger $30^{\circ}\times 30^{\circ}$ region and subsequently extracted the resulting fluxes within the central $20^{\circ}\times 20^{\circ}$ area.

For the region including the Galactic center, the emission of Fermi bubbles and the Galactic center
excess (GCE), which may have different origin from the CR interactions, need to be subtracted. 
For the Fermi bubbles, we employ the spectrum derived for $|b|>10^{\circ}$ region \cite{Fermi-LAT:2014sfa},
and re-scale to our ROI of $|b|\le 10^{\circ}$, $-10^{\circ}\le l \le 10^{\circ}$ according to the 
bubble template given in Ref.~\cite{Su:2010qj}, assuming uniform emission of the bubbles. As for the 
GCE, we adopt the results given in Ref.~\cite{Fermi-LAT:2017opo}, and again apply a re-scaling according
to the spatial profile of the GCE. The fluxes of Fermi bubbles and GCE are over-plotted in the 
central-most panel of Figure~\ref{fig:flux_componment}.

\section{Two-component model fitting of the spectra}
\label{sec:two-component}
The $\gamma$-ray spectrum from inelastic hadronic interactions between high-energy
CR protons and the ISM (assumed to be pure hydrogen) can be derived as
\begin{align}
\label{form:ppgamma}
W_{\gamma}(E_{\gamma}) = &  \int d E_{p} v_{p} n_{\rm H} \cdot
\frac{d \sigma_{p+\mathrm{H} \rightarrow \gamma}\left(E_{p}, E_{\gamma}\right)}{d E_{\gamma}}
%+n_{\rm He}\frac{d \sigma_{p+\mathrm{He} \rightarrow \gamma}\left(p_{p}, p_{\gamma}\right)}{d p_{\gamma}} 
\cdot \frac{4\pi\psi_{p}\left(E_{p}\right)}{c},
\end{align}
where $E_p$ and $E_{\gamma}$ are the (kinetic) energy of CR protons and secondary $\gamma$-ray photons,
$v_p$ is the velocity of protons and $c$ is the speed of light, $n_{\rm H}$ is the number density 
of the ISM, $d\sigma_{i+\mathrm{H} \rightarrow \gamma}/dE_{\gamma}$ is the differential cross 
section of $\gamma$-ray production, and $\psi_p$ is the flux of protons. For the cross section, 
we use the interpolation routine {\tt AAfrag} based on simulation generated
tables \cite{Kachelriess:2019ifk}.

We use two components of CRs to fit the diffuse $\gamma$-ray emission, as
\begin{align}
    \psi_p(E_p) = \psi_{\rm A}(E_p) + \psi_{\rm B}(E_p).
\end{align}
For component A, we use an interpolation of the direct measurements of proton fluxes by 
Voyager \cite{Cummings:2016pdr}, AMS-02 \cite{AMS:2021nhj}, and DAMPE \cite{DAMPE:2019gys}. 
The AMS-02 data are de-modulated to correct the solar modulation effect, based on a force-field 
model \cite{Gleeson:1968zza} with modulation potential of 0.43 GV. See the left panel of Figure
\ref{fig:proton_gamma} for the measurements and the interpolated flux of CR protons. Since the 
high-energy spectrum of protons deviate from the power-law form, which may be due to the 
contribution from a nearby source \cite{Savchenko:2015dha,Ahlers:2016njd,Liu:2018fjy}, the 
interpolation is done based on the data below 200 GeV. The $\gamma$-ray spectrum derived by 
this component is shown in the right panel of Figure \ref{fig:proton_gamma}.

The spectrum of component B, $\psi_{\rm B}$, is assumed to be a power-law with a low-energy cutoff 
\begin{align}
    \psi_{\rm B}(E_p) \propto E_p^{-\beta}\exp(-E_{\rm cut}/E_p),
\end{align}
where $\beta$ is the spectral index, and $E_{\rm cut}$ is the cutoff energy.

When fitting to the data, the normalizations of component A and B in each ROI are left free,
while the index and cutoff energy of component B are assumed to be uniform among all ROIs. 
Fitting results of the spectra in the 18 ROIs are shown in Figure \ref{fig:flux_componment}.

In Figure~\ref{fig:excess}, we plot the two-dimensional residual maps for energies
higher than 10 GeV, which are expected to be mainly from component B. To obtain the residual maps,
we subtract component A using its best-fitting spectrum times the spatial distribution (as given by
the model {\tt gll\_iem\_v07.fits}) at 1 GeV. The maps are smoothed with Gaussian kernels with
a width of 0.16 degrees. This plot reflects approximately the spatial distributions of component B
in the sky.

\newpage
\setcounter{figure}{0}
\renewcommand\thefigure{A\arabic{figure}}
\setcounter{table}{0}
\renewcommand\thetable{A\arabic{table}}

%****************************** Extended Figure 1*************************
\begin{figure*}
\centering
\centering
\includegraphics[width=0.45\linewidth]{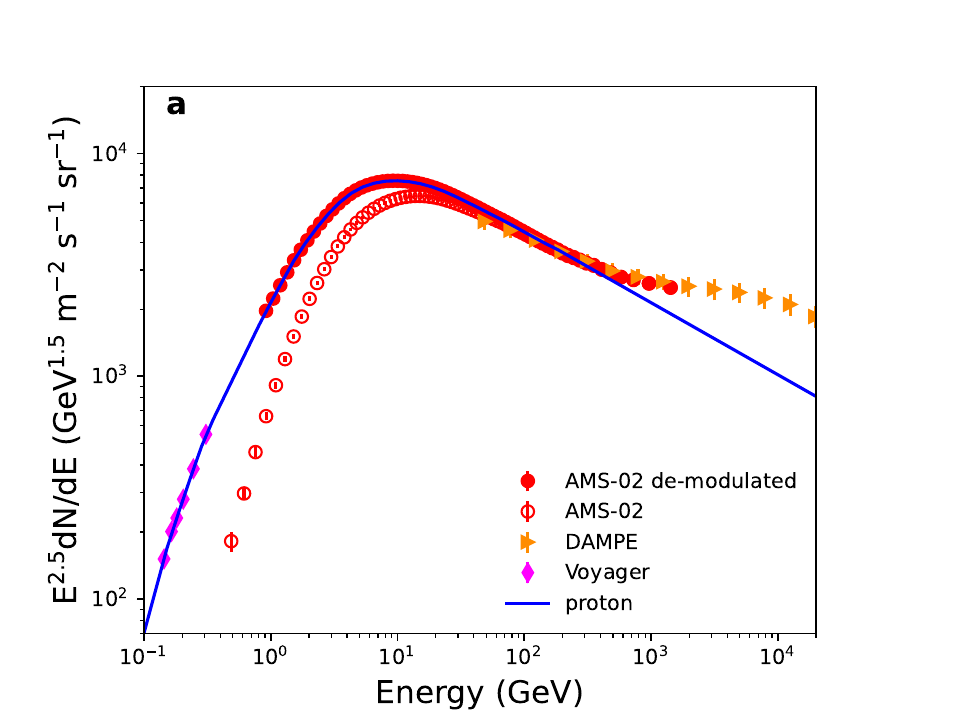}
\includegraphics[width=0.45\linewidth]{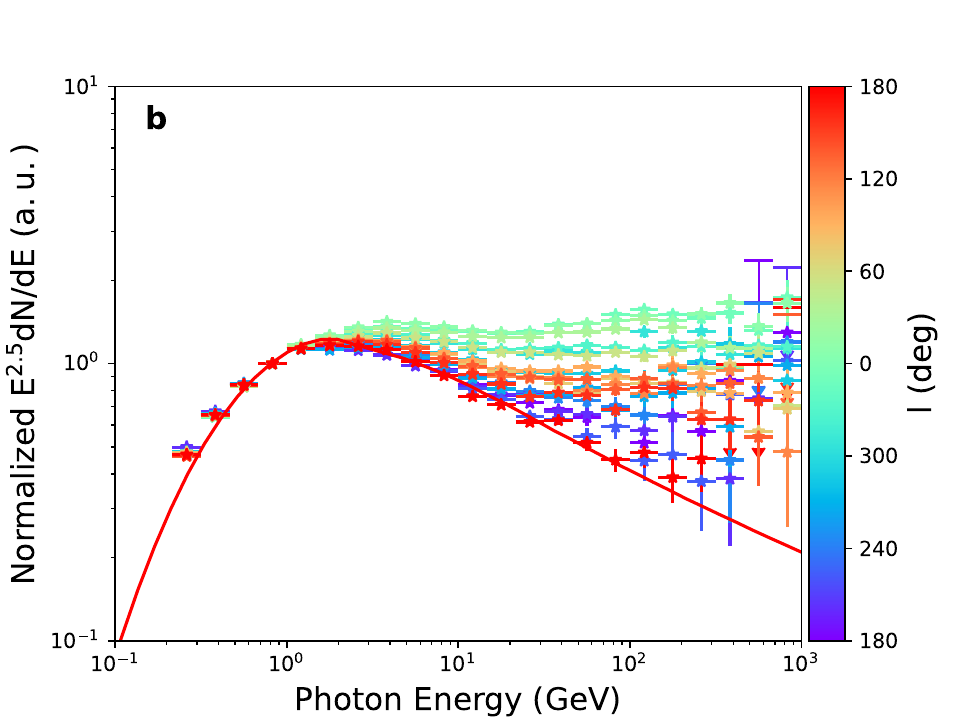}
\caption{The proton spectrum (a) and corresponding $\gamma$-ray spectrum (b) for component A.
In panel (a), the measurements by Voyager \cite{Cummings:2016pdr}, AMS-02 \cite{AMS:2021nhj}, and 
DAMPE \cite{DAMPE:2019gys} are shown. To get the proton fluxes in the local interstellar medium,
the AMS-02 data are de-modulated using the force-field model with modulation potential of 0.43 GV.
The blue line is the interpolation of the Voyager and de-modulated AMS-02 data below 200 GeV. 
Panel (b) shows the corresponding $\gamma$-ray spectrum calculated using the interpolated proton 
spectrum of panel (a), compared with the normalized measurements derived in this work.
}
\label{fig:proton_gamma}
\end{figure*}
%*************************************************************************

%****************************** Extended Figure 2*************************
\begin{figure*}
\centering
\centering
\includegraphics[width=1.0\linewidth]{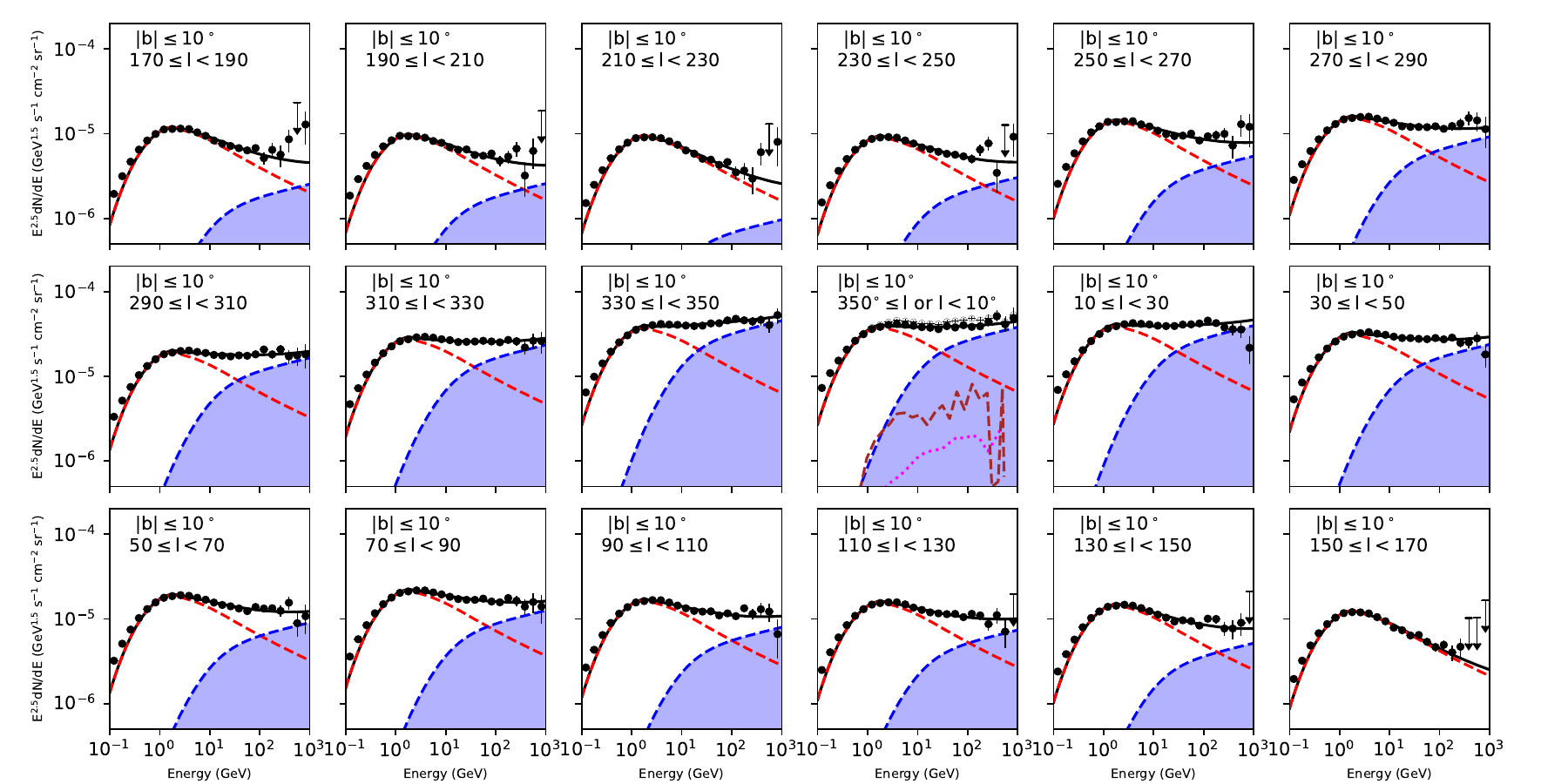}
\caption{The two-component fitting results of the SEDs in the 18 ROIs.
The black dots represent the data we analysed. The red, blue, and black lines represent the 
$\gamma$-ray spectra for component A, B, and their sum, respectively. In the Galactic center 
region, the magenta dotted line represents the spectrum of Fermi bubbles, and the brown dashed 
line represents the spectrum of the GCE. The open dots are the data including Fermi bubbles and 
the GCE, while the solid ones are fluxes excluding them.
}
\label{fig:flux_componment}
\end{figure*}
%*************************************************************************

%****************************** Extended Figure 3*************************
\begin{figure*}
\centering
\centering
\includegraphics[width=1.0\linewidth]{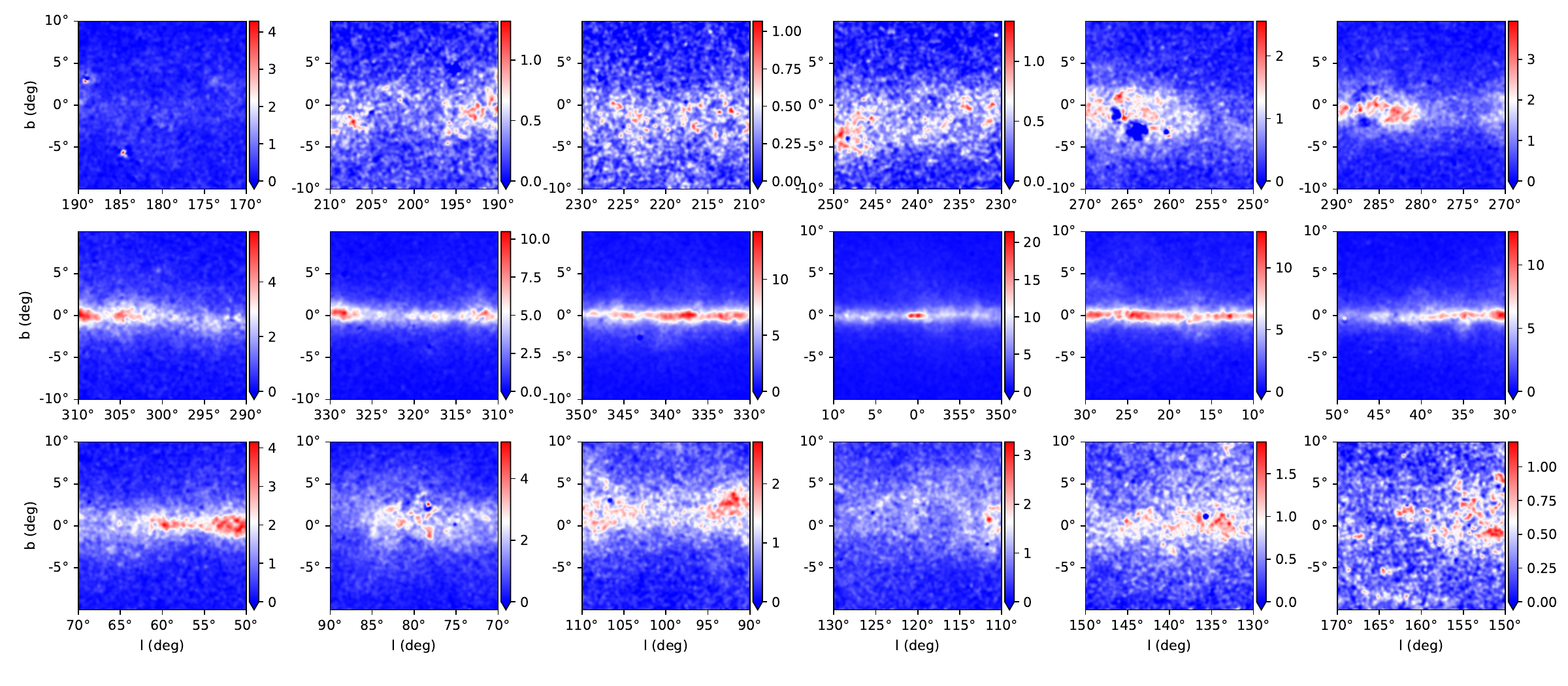}
\caption{PSF smoothed residual maps of component B in different ROIs for $E>10$ GeV. The colorbar represent the residual photons of different regions.}
\label{fig:excess}
\end{figure*}
\bibliography{ref}

\end{document}